\documentclass[12pt,a4paper]{article}

\usepackage{afterpage}
\usepackage{alltt}
\usepackage{amsmath}
\usepackage{amssymb}
\usepackage{array}
\usepackage[british]{babel}
\usepackage{booktabs}
\usepackage{caption}
\usepackage{dcolumn}
\usepackage{enumitem}
\usepackage{epstopdf}
\usepackage{eurosym}
\usepackage{fancyhdr}
\usepackage{float}
\usepackage[T1]{fontenc}
\usepackage[a4paper,margin=1in]{geometry}
\usepackage{graphicx}
\usepackage[colorlinks=true]{hyperref}
\usepackage[utf8]{inputenc}
\usepackage{latexsym}
\usepackage{longtable}
\usepackage{lscape}
\usepackage{makecell}
\usepackage{makeidx}
\usepackage{mathtools}
\usepackage[expansion=true,protrusion=true]{microtype}
\usepackage{multirow}
\usepackage{natbib}
\usepackage{newpxmath}
\usepackage{newpxtext}
\usepackage{parskip}
\usepackage{rotating}
\usepackage{setspace}
\usepackage{subcaption}
\usepackage{tabularx}
\usepackage{threeparttable}
\usepackage{tikz}
\usepackage{titlesec}
\usepackage{xcolor}

\usetikzlibrary{
  arrows.meta,
  backgrounds,
  calc,
  decorations.pathreplacing,
  fit,
  patterns,
  positioning,
  shapes.geometric,
  shapes.misc
}

\tikzset{
  block/.style={rectangle,draw=black, thick,rounded corners,align=center,minimum height=9mm,minimum width=31mm,inner sep=4pt},
  smallblock/.style={rectangle,draw=black,thick,rounded corners,align=center,minimum height=7mm,minimum width=24mm,inner sep=3pt},
  box/.style={draw,rounded corners=3pt,align=center,minimum height=10mm,minimum width=25mm,inner sep=4pt},
  smallbox/.style={draw,rounded corners=3pt,align=center,minimum height=8mm,minimum width=18mm,inner sep=3pt},
  flow/.style={->, line width=0.9pt},
  group/.style={draw, rounded corners=4pt, inner sep=6pt},
  axislabel/.style={font=\small},
  titlelabel/.style={font=\small},
  state/.style={ellipse,draw=black, thick,align=center,minimum height=9mm,minimum width=27mm,inner sep=3pt},
  policy/.style={rectangle,draw=black, thick,align=center,minimum height=8mm,minimum width=26mm,inner sep=3pt},
  nonlinear/.style={rectangle,draw=black,dashed,thick,rounded corners,align=center,minimum height=8mm,minimum width=30mm,inner sep=3pt},
  arrow/.style={-{Latex[length=3mm]}, thick},
  dashedarrow/.style={-{Latex[length=3mm]}, thick,dashed},
  feedback/.style={-{Latex[length=3mm]}, thick,bend left=25},
  thinarr/.style={-{Latex[length=2.2mm]},thick},
  note/.style={align=center,font=\small}
}

\bibpunct{(}{)}{;}{a}{,}{,}
\hypersetup{linkcolor=black,urlcolor=black,citecolor=black}
\setlist{noitemsep, topsep=0pt}

\title{Geoeconomic Stress and \\European Defence Readiness%
\thanks{The authors acknowledge helpful comments from participants of meetings of the NATO STO Operations Research and Analysis conferences, and NATO STO Modelling and Simulation Group meetings. The authors are solely responsible for the content of the paper. The views expressed are purely those of the authors and may not in any circumstances be regarded as stating an official position of the European Commission, NATO or Latvian National Armed Forces. Any remaining errors are ours.}}
\author{d'Artis Kancs%
\thanks{Science Research Innovation Implementation Centre, Latvian National Armed Forces.}}
\date{12 January 2026}

\begin{document}

\maketitle

\begin{abstract}
This study develops a country-level measure of the economic-industrial capacity of European NATO Allies to mobilise, replenish, maintain, and sustain military capability under strategic stress. The measure comprises four components: macroeconomic capacity, sectoral resilience, defence-industrial capability, and logistics and sustainment. The analysis covers 28 European Allies and evaluates a balanced baseline against five geoeconomic stress scenarios. The results distinguish baseline capacity from stress vulnerability and provide an assessment framework for identifying the economic-industrial bottlenecks that can constrain sustained military readiness. Baseline scores vary materially across countries, with a sample mean of 0.610 and a range from 0.471 to 0.759. The calibrated geoeconomic shocks reduce effective capacity in every scenario, though the effects are highly heterogeneous. The combined-stress scenario has the largest average effect, lowering the sample mean to 0.491, a decline of 19.5 per cent relative to baseline. TRS and PR operate mainly through defence-industrial and replenishment constraints; supply-chain fragmentation is driven primarily by sectoral exposure and substitutability; and external-support reduction operates through allied access and sustainment. Our results suggest shifting attention from expenditure targets alone towards the capacity to convert resources into usable military capability: the key issue is not only how much is spent, but also which bottlenecks limit the conversion of resources into capability.
\end{abstract}

\noindent\textbf{Keywords:} Defence economics; readiness; geoeconomic fragmentation; supply chains; defence-industrial base; Europe; strategic stockpiling; procurement.

\noindent\textbf{JEL Codes:} F51; F52; H56; L64; O25; R11.

\clearpage

\section{Introduction}
\label{sec:introduction}

Europe's security environment has entered a period of sustained strategic stress. Russia's full-scale invasion of Ukraine has increased the salience of conventional deterrence and defence, while competition in other theatres can absorb allied production capacity, munitions, air defence, intelligence assets, sealift, maintenance resources, and political attention \citep{kendall-taylor2025,lee2024,cancianpark2026}. At the same time, trade, finance, technology transfer, raw-material access, energy, shipping, and industrial capacity are increasingly shaped by geopolitical alignment \citep{farrell2019,aiyar2023,europeancommission2024edis,nato2024capacitypledge}. These developments raise a question that defence expenditure alone cannot answer: can European states convert available resources into military capability quickly enough, and sustain that conversion when economic and industrial conditions deteriorate?

This study asks how geoeconomic and industrial conditions shape the capacity of European NATO Allies to mobilise, replenish, and sustain military capability under prolonged strategic stress. The analysis focuses on \emph{readiness-enabling conversion capacity}: the capacity to convert economic, industrial, and logistical resources into replacement, replenishment, maintenance, and sustained military capability. This capacity is primarily a dimension of mobilisation readiness and is analytically distinct from both general economic resilience \citep{alloush2024} and contemporaneous operational readiness \citep{betts1995}.

We make three contributions. First, we conceptualise and operationalise the intermediate conversion mechanism linking economic resources to sustained military capability. Second, we develop a transparent country-level measurement framework that integrates macroeconomic capacity (\(M\)), defence-relevant supply-chain resilience (\(S\)), defence-industrial capability (\(D\)), logistics and sustainment (\(L\)), and effective access to allied or foreign supply. Third, we separate baseline capacity from stress vulnerability by applying harmonised geoeconomic shocks within the same measurement framework, making it possible to assess heterogeneous effects on countries that start from similar readiness levels but respond differently when supply, replenishment, or external support deteriorates.

The empirical analysis covers 28 European NATO Allies. The composite index combines four components---\(M\), \(S\), \(D\), and \(L\)---constructed from 25 indicators. A balanced baseline is stress-tested against five scenarios: transfer-and-replenishment stress, prolonged rivalry, supply-chain fragmentation, external-support reduction, and combined stress. Low, Central, and High calibrations capture scenario intensity. The analysis is conducted at the country level, and results are decomposed to component and indicator level. Independent external benchmarks are used to assess whether the resulting index exhibits the expected relationships with industrial, innovation, and logistics capabilities.

The empirical analysis is a structured measurement and stress-testing exercise rather than an estimated causal model or conflict forecast. Scenario results identify how Allies' conversion capacity changes under common calibrated geoeconomic shocks; they do not estimate battlefield outcomes or the probability of military success. This distinction is important because actual crises are heterogeneous and can expose countries to very different levels of physical destruction, attrition, infrastructure disruption, and external support, as Russia's full-scale invasion of Ukraine demonstrates.

The remainder of the paper proceeds as follows. Section 2 develops the conceptual framework and empirical propositions. Section 3 compares the measure with existing readiness, mobilisation, defence-industrial, resilience, and geoeconomic approaches. Section 4 describes the sample, data architecture, and scenario design. Section 5 presents the index. Section 6 reports the empirical results. Section 7 presents external validation. Section 8 discusses interpretation, policy implications, and limitations. Section 9 concludes. The online appendix documents the full indicator construction, coverage matrix, scoring methodology, scenario calibration, transformation equations, and robustness procedures.

\section{Conceptual framework}
\label{sec:readiness_theory}
\label{sec:framework}

\subsection{Readiness concepts and the analytical boundary}

Military readiness is not determined by military inventories alone. In a protracted crisis or conflict, a state must also be able to replace equipment, replenish stocks, maintain platforms, expand production, secure critical inputs, and sustain the movement of materiel. These functions draw on economic and industrial resources that lie partly outside the armed forces. The analytical question is therefore not only what resources a country possesses, but how effectively they can be converted into usable and sustainable military capability when conditions deteriorate.

Following \citet{betts1995}, the framework distinguishes three related concepts. \emph{Structural readiness} concerns the forces, equipment, and force structure available for employment. \emph{Operational readiness} concerns the ability of existing forces to perform assigned missions. \emph{Mobilisation readiness} concerns the capacity to generate, expand, replace, replenish, and sustain capability beyond the forces and stocks already available. These concepts are connected but not interchangeable. A force can be operationally ready at the beginning of a conflict yet possess limited capacity to replace losses or replenish munitions. Conversely, a state can possess substantial industrial potential without maintaining a powerful standing force.

The analysis focuses primarily on mobilisation readiness. We define \emph{readiness-enabling conversion capacity} as the capacity of a state to convert economic, industrial, and logistical resources into replacement, replenishment, maintenance, and sustained military capability under strategic stress. This capacity is narrower than national preparedness or whole-of-society resilience, but broader than defence-industrial production alone. It captures the economic-industrial conditions that enable military capability to be regenerated and sustained once initial stocks, peacetime production rates, and established supply arrangements come under pressure.

This formulation builds on the distinction between possessing latent resources and being able to translate them into military power. \citet{kancs2025} characterises mobilisation readiness in terms of the convertibility and expansibility of civilian personnel and industrial resources and similarly links latent power, defence investment, capabilities, capacity, and security outcomes.

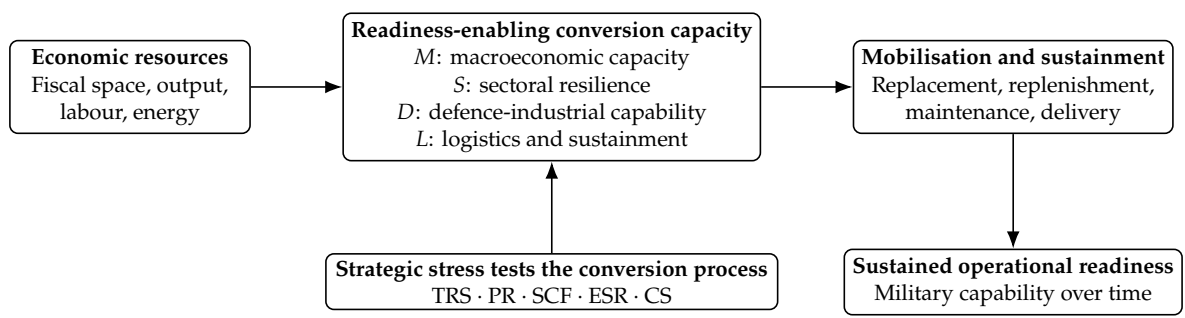
\begin{figure}[htbp]
\centering
\resizebox{0.98\textwidth}{!}{%
\begin{tikzpicture}[node distance=12mm and 16mm, font=\small]
\node[block, minimum width=42mm] (resources) {\textbf{Economic resources}\\Fiscal space, output,\\labour, energy};
\node[block, right=of resources, minimum width=56mm] (conversion) {\textbf{Readiness-enabling conversion capacity}\\
\(M\): macroeconomic capacity\\
\(S\): sectoral resilience\\
\(D\): defence-industrial capability\\
\(L\): logistics and sustainment};
\node[block, right=of conversion, minimum width=45mm] (sustain) {\textbf{Mobilisation and sustainment}\\Replacement, replenishment,\\maintenance, delivery};
\node[block, below=20mm of sustain, minimum width=42mm] (operational) {\textbf{Sustained operational readiness}\\Military capability over time};

\draw[arrow] (resources) -- (conversion);
\draw[arrow] (conversion) -- (sustain);
\draw[arrow] (sustain) -- (operational);

\node[smallblock, below=16mm of conversion, minimum width=79mm] (stress) {\textbf{Strategic stress tests the conversion process}\\
TRS \(\cdot\) PR \(\cdot\) SCF \(\cdot\) ESR \(\cdot\) CS};
\draw[arrow] (stress.north) -- (conversion.south);
\end{tikzpicture}
}
\caption{Readiness-enabling conversion capacity in the readiness chain}
\label{fig:conversion_capacity_framework}
\end{figure}

Readiness-enabling conversion capacity therefore occupies an intermediate position between general economic resilience and sustained operational readiness (Figure~\ref{fig:conversion_capacity_framework}). Economic resilience preserves the resource base from which mobilisation can occur; conversion capacity determines whether those resources can be directed into defence production, replacement, maintenance, and sustainment; operational readiness concerns whether forces can perform assigned missions. We do not measure combat effectiveness, force quality, or contemporaneous operational readiness. A high index value indicates favourable economic-industrial conditions for sustaining capability under stress, not that a country's armed forces would prevail in a particular military operation.

\subsection{Four mechanisms of conversion capacity}

The empirical framework decomposes readiness-enabling conversion capacity into four mechanisms: macroeconomic capacity \(M\), sectoral resilience \(S\), defence-industrial capability \(D\), and logistics and sustainment \(L\). The decomposition separates resource availability from the specific mechanisms through which those resources become militarily useful.

The four-part decomposition follows the economic sequence from resource mobilisation, through input continuity and defence-specific production, to sustainment and delivery. Mobilisation research motivates the aggregate resource constraint \citep{harrison1988}; production-network and geoeconomic research identifies direct and indirect input dependencies \citep{acemoglu2012,farrell2019,kancs2024supplychains}; defence-economics and mobilisation research highlights procurement, surge capacity, and lead times \citep{smith2022,mueller2025,kancs2025}; and military-logistics research links maintenance, transport, and repair capacity to operational availability \citep{mcgee2005,garciabenito2021}.

\paragraph{Macroeconomic capacity \(M\).}
Government fiscal space, economic scale, domestic absorption capacity, inflationary pressures, energy dependence, and labour availability condition the feasible rate at which defence demand can expand. Mobilisation entails opportunity costs because labour, finance, energy, and productive capacity have alternative civilian uses. Macroeconomic capacity therefore defines the resource envelope within which industrial mobilisation occurs. It is a necessary, though not sufficient, condition: a large economy with weak defence-specific production or severe supply dependencies may convert resources more slowly than a smaller but better-prepared industrial system \citep{harrison1988,kancs2025}.

\paragraph{Sectoral resilience \(S\).}
Advanced defence production relies on multi-tier supply chains containing specialised components, critical raw materials, machine tools, electronics, and technologies for which substitution can require lengthy qualification or redesign. Direct import dependence captures only first-order exposure. Production can also depend indirectly on foreign inputs embodied in domestically purchased intermediates. Sectoral resilience therefore depends on direct and indirect exposure, supplier concentration, the availability of alternatives, and the technical and logistical feasibility of substitution. Security of supply becomes a constraint on nominal industrial capacity when suppliers restrict exports, transport deteriorates, or simultaneous allied demand rises \citep{acemoglu2012,kancs2024supplychains}.

\paragraph{Defence-industrial capability \(D\).}
The defence-industrial mechanism determines whether demand can be translated into materiel at the speed of relevance. Supplier depth, baseline production, usable surge capacity, procurement activity, qualification bottlenecks, lead times, and access to foreign or allied supply all affect effective output \citep{cancian2020}. The relevant distinction is between nominal and expandable capacity. Measured headroom is not necessarily immediately usable because additional production can require specialised labour, energetic materials, components, tooling, testing facilities, or regulatory qualification. Time therefore matters directly: a production system capable of eventually replacing a platform is not equivalent to one able to replace it within the operationally relevant period. \citet{kancs2025} makes this time dimension explicit through inventory-expansion time, which depends on the required inventory, production rate, and production lead time.

\paragraph{Logistics and sustainment \(L\).}
Conversion is incomplete until generated capability can be maintained and delivered. Transport infrastructure, maintenance and repair capacity, spare-parts availability, logistics-support capacity, replenishment speed, and replacement flows determine whether military systems remain available after initial use. Initial stocks can temporarily compensate for slow production or delivery, but their contribution declines as equipment requires repair and inventories are consumed. Logistics and sustainment therefore connect industrial output to persistent operational capability \citep{mcgee2005,garciabenito2021}.

The four mechanisms also define the boundary with linked concepts of readiness. Force structure and contemporaneous operational readiness are downstream outcomes rather than conversion mechanisms; broad societal resilience is upstream and wider than the capacity we study; defence procurement and defence-specific labour constraints enter \(D\); stocks, maintenance, transport, and replenishment enter \(L\); and allied or foreign supply access enters the production and sustainment mechanisms rather than forming a separate fifth capacity. This nesting keeps the framework parsimonious while retaining the principal economic-industrial constraints identified in the literature \citep{betts1995,kancs2025,smith2022,mcgee2005}.

\subsection{Strategic stress, allied access, and vulnerability}

Peacetime capacity and stressed capacity are not the same. Many arrangements that increase efficiency in normal conditions---international specialisation, lean inventories, concentrated suppliers, or reliance on allied production---can create contingent vulnerability when access, transport, or simultaneous demand changes.

Domestic production is only one source of replenishment. For capability or production category \(g\), effective supply is represented as:
\begin{equation}
Q^{E}_{c,g,s}
=
Q^{DOM}_{c,g,s}
+
A_{c,g,s}Q^{FOR}_{c,g,s},
\label{eq:effective_replenishment_supply}
\end{equation}
where \(Q^{DOM}\) is effective domestic supply, \(Q^{FOR}\) is potentially available foreign or allied supply, and \(A_{c,g,s}\in[0,1]\) is the effective foreign-access factor. \(A\) captures contractual access, expected delivery availability, supplier-country concentration, export and licensing restrictions, transport conditions, competing allied demand, and emergency access arrangements. This formulation treats allied supply as part of effective capacity rather than equating readiness with domestic self-sufficiency.

Strategic stress acts on the conversion process through several channels. Trade fragmentation and technology denial affect imported inputs and substitutability; transfer and replenishment pressure raises replacement demand and production queues; prolonged rivalry changes the duration over which production and sustainment constraints operate; and reduced external support lowers effective foreign access. These shocks can propagate from macroeconomic and sectoral exposure into production bottlenecks, longer lead times, maintenance pressure, spare-parts scarcity, and slower replenishment. A narrow disruption to a defence-critical input can therefore have a small effect on aggregate GDP but a large effect on military conversion capacity, while a broad trade shock can have a modest readiness effect if the affected inputs are readily substitutable or militarily unimportant.

For country \(c\), let baseline conversion capacity be \(R_{c,0}\) and stressed capacity under scenario \(s\) be \(R_{c,s}\). Scenario vulnerability is:
\begin{equation}
V_{c,s}
=
R_{c,0}-R_{c,s}.
\label{eq:theory_vulnerability}
\end{equation}
Baseline capacity and stress vulnerability therefore capture distinct dimensions of readiness. Countries with similar baseline scores can experience heterogeneous losses under the same scenario because their capacity rests on different combinations of domestic production, imported inputs, supplier diversity, allied access, and logistics.

\subsection{Complementarity, bottlenecks, and time}

Decomposing readiness-enabling conversion capacity into four mechanisms helps identify bottlenecks and clarify how they interact. In operational and technological terms, however, the mechanisms are complementary. Fiscal resources cannot replace a missing qualified component immediately. Production capacity has limited military value if maintenance and delivery systems cannot sustain the resulting equipment. Allied procurement cannot fully compensate for domestic shortfalls when allies face simultaneous replenishment requirements. Infrastructure cannot substitute for unavailable replacement materiel.

Readiness-enabling conversion capacity therefore exhibits a bottleneck property, whereby system output may be disproportionately constrained by its weakest component. This does not require the empirical index to impose a strict Leontief production function. It implies instead that stress can reveal dependencies that remain latent in normal conditions. The component decomposition is therefore provided alongside the composite index because similar aggregate scores can reflect very different configurations of constraints.

Time introduces a second distinction between \emph{stock} and \emph{flow} constraints. Stock constraints concern inventories, stored munitions, spare parts, and pre-positioned equipment. Flow constraints concern the speed of production, contracting, maintenance, transport, and replacement. A country with adequate initial stocks but weak replenishment can appear resilient in the first phase of a crisis and constrained later; a country with limited stocks but stronger replenishment can face immediate vulnerability yet regenerate more rapidly. The five stress scenarios are designed to expose these differences rather than to forecast a single conflict path.

The mechanism is sequential for analytical clarity but allows feedback: missing inputs can delay production and accelerate inventory drawdown; logistics problems can increase emergency procurement demand; and external-support diversion can redirect demand towards already constrained domestic suppliers.

\subsection{Empirical propositions}

The conceptual framework yields four propositions about the relationships that should be visible in the country-level decomposition and stress tests.

\paragraph{Proposition 1: Conversion and bottlenecks.}
Readiness-enabling conversion capacity depends jointly on macroeconomic resources, resilient strategic inputs, defence-industrial capability, and logistics and sustainment. Because these mechanisms are complementary, a weak component can constrain effective conversion even when other components are strong:
\begin{equation}
R_c=f(M_c,S_c,D_c,L_c),
\qquad
\frac{\partial R}{\partial M},
\frac{\partial R}{\partial S},
\frac{\partial R}{\partial D},
\frac{\partial R}{\partial L}>0.
\label{eq:prop_conversion}
\end{equation}

\paragraph{Proposition 2: Stress heterogeneity.}
Geoeconomic shocks operating through different channels generate different patterns of country vulnerability. Replenishment pressure should load primarily on defence-industrial and sustainment mechanisms, while supply-chain fragmentation should load more strongly on sectoral exposure, concentration, and substitutability. Distinct channel contributions and country vulnerability profiles therefore provide evidence that the scenarios reveal different constraints rather than reproducing a single generic stress ranking.

\paragraph{Proposition 3: Allied-access contingency.}
Allied and foreign supply expands effective conversion capacity when access is reliable, but dependence on that supply creates contingent vulnerability when contractual access, delivery, transport, or allied availability deteriorates:
\begin{equation}
\frac{\partial V_{c,ESR}}
{\partial\,\text{foreign-access dependence}}
>0.
\label{eq:prop_allied_access}
\end{equation}

\paragraph{Proposition 4: Compound stress.}
When several complementary constraints deteriorate simultaneously, combined stress should produce a larger loss of effective conversion capacity than the corresponding single-channel stresses, subject to the model's non-double-counting rules:
\begin{equation}
V_{c,CS}
\geq
\max
\left(
V_{c,TRS},
V_{c,SCF},
V_{c,ESR}
\right).
\label{eq:prop_combined}
\end{equation}

\section{Related literature}
\label{sec:literature_policy_context}

Our key contribution is best understood through a measurement comparison: \emph{what do existing approaches measure, and what remains missing?} The relevant strands of literature provide strong treatments of military readiness, mobilisation, defence-industrial capacity, societal resilience, and geoeconomic dependence, but they generally observe different parts of the readiness chain. The gap is not an absence of readiness indicators. It is the absence of a transparent, country-comparable measure of the economic-industrial conversion mechanism defined in Section~\ref{sec:readiness_theory}.

\subsection{Military readiness and mobilisation}

Military-readiness systems are primarily designed to assess the condition and deployability of forces. They organise information around personnel, equipment, training, supply, maintenance, and mission capability \citep{betts1995,dodi773066,usmc2020readiness}. Their unit of analysis is commonly a force, service, unit, or mission, and many of the most decision-relevant measures are classified or organisation-specific. These systems are therefore closely linked to military decision-making but are not designed to compare states’ economic-industrial capacity to regenerate capability after stocks are depleted or equipment is damaged.

Mobilisation research is more closely aligned with the present study because it focuses on expansion and regeneration. \citet{kancs2025}, for example, measures mobilisation through the convertibility and expansibility of personnel, materiel, and industrial resources. Other mobilisation approaches use inventories, production lines, mobilisation plans, historical benchmarks, or capability-specific replacement requirements. These measures reveal important bottlenecks but are often tied to particular systems, national plans, or restricted information. They do not generally provide a harmonised country-level measure of the broader conversion mechanism.

The present study therefore complements the existing structural and operational readiness measures. Our empirical object is upstream of mission readiness: the economic-industrial capacity that enables mobilisation, replenishment, maintenance, and sustained force availability once peacetime stocks and production arrangements come under pressure.

\subsection{Defence-industrial capacity and mobilisation}

Defence economics has long emphasised features that distinguish defence markets from ordinary product markets: state-dominated demand, security restrictions, low-volume specialised production, long development cycles, scale economies, uncertainty, and the strategic value of retaining production capabilities that often remain underused for longer periods in peacetime \citep{sandlerhartley1995,dunne1995,hartley2007}. Work on defence output and procurement further distinguishes budgetary inputs from delivered military capability \citep{hartley2012,iiss2022}. Recent policy debates and academic literature place greater emphasis on surge capacity, production bottlenecks, procurement credibility, supply-chain resilience, and sustained replenishment demand \citep{cancian2020,nato2024capacitypledge,nato2025dpap,europeancommission2024edis}.

This literature identifies a central part of the conversion mechanism. Defence expenditure produces military output only if demand can pass through contracting, qualified suppliers, specialised labour, production capacity, certification, testing, maintenance, and delivery. Large increases in budgets or orders therefore need not translate into proportionate near-term output. \citet{kancs2025} makes the time constraint explicit through inventory-expansion time, while analyses of munitions and replenishment similarly emphasise the speed at which stocks can be regenerated \citep{cancianpark2026}.

The measurement gap is one of scope. Defence-industrial studies generally examine particular weapon systems, production lines, firms, procurement institutions, or national industrial bases. Industrial capacity is indispensable to the present analysis, but it does not by itself capture macroeconomic absorption, indirect geoeconomic exposure, logistics and maintenance, or contingent access to allied supply. The \(D\) component is therefore one mechanism within a broader conversion system.

\subsection{Economic and societal resilience}

A third strand of literature studies to what extent economies and societies can absorb shocks, maintain essential functions, adapt to disruption, and recover \citep{alloush2024}. In defence policy this includes continuity of government and critical services, infrastructure, civil preparedness, stockpiles, production and distribution systems, and the capacity of society to sustain essential functions during a protracted crisis. \citet{kancs2025}, for example, treats sustained whole-of-society resilience as a strategic-readiness dimension and compares preparedness, shock resistance, and recovery capacity across countries.

This literature is closely related to the paper's resource base but measures a broader object. Resilience literature asks whether economic and societal functions persist under disruption \citep{alloush2024}. It does not normally ask whether those resources can be converted into additional ammunition, repaired platforms, qualified substitute inputs, spare parts, maintenance capacity, or military logistics at the required rate. A country can maintain aggregate output and essential services while still facing a narrow defence-industrial bottleneck; conversely, a broad macroeconomic shock need not become a military bottleneck if defence-critical inputs and conversion channels remain available.

The distinction is therefore at the analytical level. Economic and societal resilience is conceptually upstream of readiness-enabling conversion capacity. External validation can reveal whether the concepts overlap where expected without requiring a defence-specific conversion measure to reproduce a general resilience index.

\subsection{Geoeconomic vulnerability and supply-chain resilience}

Research on global value chains, production networks, and geoeconomic fragmentation shows that final production depends on international chains of intermediate inputs and that shocks propagate beyond directly exposed firms and sectors \citep{hummels2001,antras2013,acemoglu2012,boehm2019,bonadio2021,baqaee2024}. The geoeconomic literature adds a strategic dimension: networks that improve efficiency can also create leverage and vulnerability when access to trade, finance, technology, transport nodes, or critical raw materials becomes politically constrained \citep{farrell2019,aiyar2023}. Work on critical raw materials further shows that exposure depends on concentrated extraction and processing, by-product dependence, and the difficulty of expanding supply rapidly \citep{nassar2015,graedel2015,oecd2023,europeancommission2023}.

This literature supplies the transmission logic for the sectoral-resilience component. Direct import dependence captures first-order exposure; input-output linkages capture foreign inputs embedded in domestically purchased intermediates; supplier concentration affects the availability of alternatives; and qualification or input specificity determines whether nominal substitutes can be used within the relevant horizon.

The missing link is military conversion. Geoeconomic vulnerability is commonly measured in terms of trade exposure, output, welfare, or dependence on foreign inputs and markets. A large aggregate exposure can have limited readiness consequences when inputs are readily substitutable or militarily unimportant, while a small trade flow can become strategically important when it contains a certified component with no near-term substitute. The present framework therefore maps geoeconomic exposure into defence-relevant production and sustainment by combining direct and indirect dependence with concentration, substitutability, critical-input relevance, production lead times, and foreign-access conditions.

\subsection{Composite measurement}

Taken together, existing studies identify the relevant mechanisms that condition sustained military capability, but they measure different objects. Military-readiness systems measure force condition and mission capability; mobilisation studies measure selected expansion and regeneration capacities; defence-industrial research measures production and procurement constraints; resilience indices measure shock absorption and recovery; and geoeconomic measures identify external dependencies and propagation channels. Table~\ref{tab:measurement_comparison} summarises the distinction. None of these approaches is designed to provide a harmonised, reproducible country-level measure of the intermediate conversion mechanism linking economic resources to military replenishment and sustainment across European states.

\begin{table}[!htbp]
\centering
\small
\renewcommand{\arraystretch}{1.00}
\begin{threeparttable}
\caption{Measurement comparison: existing approaches and readiness-enabling conversion capacity}
\label{tab:measurement_comparison}
{\linespread{0.96}\selectfont
\begin{tabularx}{\textwidth}{@{}
>{\raggedright\arraybackslash}p{0.15\textwidth}
>{\raggedright\arraybackslash}p{0.20\textwidth}
>{\raggedright\arraybackslash}p{0.16\textwidth}
>{\raggedright\arraybackslash}X
@{}}
\toprule
\makecell[l]{Approach} &
\makecell[l]{Primary object\\measured} &
\makecell[l]{Typical unit\\of analysis} &
\makecell[l]{Measurement gap relative\\to this paper} \\
\midrule

Military readiness &
Force condition, training, mission capability, equipment, personnel &
Unit, service, force, mission &
Measures operational deployability but generally not the economic-industrial capacity to regenerate capability under prolonged strategic stress. \\
\addlinespace[2pt]

Mobilisation and preparedness &
Expansion of personnel, materiel, inventories, and selected industrial capacity &
Country, force, capability, weapon system &
Captures regeneration directly but often through selected stocks, production lines, historical benchmarks, or restricted planning information rather than a harmonised country framework. \\
\addlinespace[2pt]

Defence-industrial capacity &
Production, procurement, supplier base, lead times, surge, industrial policy &
Firm, sector, programme, weapon system, country &
Measures the industrial conversion stage but usually does not jointly incorporate macroeconomic absorption, indirect geoeconomic exposure, logistics, sustainment, and contingent foreign access. \\
\addlinespace[2pt]

Economic and societal resilience &
Shock absorption, continuity, adaptation, and recovery &
Country, economy, society, infrastructure system &
Measures whether essential economic and societal functions persist, not whether resources can be converted into military replacement and replenishment at the required rate. \\
\addlinespace[2pt]

Geoeconomic and supply-chain vulnerability &
Trade dependence, network exposure, concentration, fragmentation, critical inputs &
Product, sector, firm, country, network &
Identifies where shocks propagate but not whether the affected dependency becomes a binding military production, maintenance, or replenishment constraint. \\
\addlinespace[2pt]

\textbf{This study} &
\textbf{Readiness-enabling conversion capacity} &
\textbf{Country, with indicator and component decomposition} &
\textbf{Links macroeconomic capacity, defence-relevant supply-chain resilience, defence-industrial capability, logistics and sustainment, and effective foreign access under common strategic stress scenarios.} \\

\bottomrule
\end{tabularx}
}
\begin{tablenotes}[flushleft]
\footnotesize
\item Note: The categories summarise dominant measurement objects rather than mutually exclusive literatures. Individual studies may span more than one category.
\end{tablenotes}
\end{threeparttable}
\end{table}

\section{Empirical implementation}
\label{sec:scenario_data}

\subsection{Country sample}

The empirical sample covers 28 European NATO Allies: Albania, Belgium, Bulgaria, Croatia, Czechia, Denmark, Estonia, Finland, France, Germany, Hungary, Iceland, Italy, Latvia, Lithuania, Luxembourg, Montenegro, the Netherlands, North Macedonia, Norway, Poland, Portugal, Romania, Slovakia, Slovenia, Spain, Sweden, and the United Kingdom. The United States and Canada are excluded because the analysis concerns European defence readiness.

Greece and Türkiye are excluded by a pre-specified comparability rule. Their defence postures are shaped to a significant degree by security requirements that are analytically distinct from the Russian conventional threat that defines the comparison set: Aegean and Eastern Mediterranean security, bilateral Greek--Turkish competition, Cyprus, wider Middle Eastern contingencies, maritime access, and other regional requirements. The exclusion does not imply that either state is unaffected by Russia or does not contribute to NATO deterrence. It reflects the fact that the same observed indicator can represent a different underlying readiness requirement across the Aegean and northern/eastern European Allies.

\subsection{Public replication and restricted validation}

The empirical architecture separates the public measurement layer from a restricted validation environment. Let \(R^{P}_{c,s}\) denote the public country-level index for country \(c\) under scenario \(s\):
\begin{equation}
R^{P}_{c,s}
=
w_{M,s}M_{c,s}
+
w_{S,s}S_{c,s}
+
w_{D,s}D^{P}_{c,s}
+
w_{L,s}L^{P}_{c,s},
\label{eq:public_core_index}
\end{equation}
where \(D^{P}\) and \(L^{P}\) use public or reproducible proxy information. Let \(R^{A}_{c,s}\) denote an augmented validation result using authorised restricted information. The validation difference is:
\begin{equation}
\Delta^{R}_{c,s}
=
R^{A}_{c,s}-R^{P}_{c,s}.
\label{eq:restricted_validation_gap}
\end{equation}
The public model produces the presented results. Restricted inventories, target stocks, maintenance states, production rates, or contract information test whether more precise information materially changes country levels, ranks, binding components, or scenario sensitivities, but they do not define a result that cannot be reproduced from the public model.
\footnote{To simplify notation, the remainder of the paper writes the readiness index as \(R_{c,s}\equiv R^{P}_{c,s}\). The superscript is retained only when the distinction between the public and augmented layers is discussed.}

\subsection{Variables and data sources}

Table~\ref{tab:data_components} summarises the four public-core components. The online appendix provides the detailed component indicator registers and scoring rules in Tables 4--7, the partial-observation register in Table 8, and the applied normalisation bounds in Table 9, including exact source identifiers, transformations, weights, vintages, country coverage, missing-data rules, and public/restricted status for the indicators used in computations.

\begin{table}[!htbp]
\centering
\small
\renewcommand{\arraystretch}{1.00}
\begin{threeparttable}
\caption{Variables and data sources: overview}
\label{tab:data_components}

{\linespread{0.96}\selectfont
\begin{tabularx}{\textwidth}{@{}
>{\raggedright\arraybackslash}p{0.08\textwidth}
>{\raggedright\arraybackslash}p{0.18\textwidth}
>{\raggedright\arraybackslash}p{0.32\textwidth}
>{\raggedright\arraybackslash}X
@{}}
\toprule
\makecell[l]{Compo-\\nent} &
\makecell[l]{Concept\\measured} &
\makecell[l]{Public-core\\indicator groups} &
\makecell[l]{Principal public\\sources} \\
\midrule

\(M\) &
Macroeconomic capacity &
Fiscal headroom, price exposure, labour/absorption capacity, energy sensitivity &
Eurostat, IMF, ECB, OECD and national statistical sources \\
\addlinespace[2pt]

\(S\) &
Sectoral resilience &
Import concentration, upstream exposure, substitutability, critical-material dependence &
OECD ICIO/TiVA, Eurostat, Comtrade/BACI, PRODCOM, JRC/EC critical-material sources \\
\addlinespace[2pt]

\(D^{P}\) &
Public defence-industrial capability &
Supplier depth, procurement activity, public production proxies, lead-time evidence, certification and surge proxies &
TED and national procurement portals, firm reports, Eurostat SBS/PRODCOM sources, EDA/NATO/EC industrial sources \\
\addlinespace[2pt]

\(L^{P}\) &
Public/proxy logistics and sustainment &
Transport and reinforcement infrastructure, public maintenance/MRO capacity, replenishment evidence, inventory-replacement proxies &
TEN-T/infrastructure data, procurement and MRO sources, public readiness and logistics reports \\

\bottomrule
\end{tabularx}
}

\begin{tablenotes}[flushleft]
\footnotesize
\item Note: The table defines Layer A, the public replication model. Restricted inventories, target stocks, maintenance states, production rates, or contract information enter only Layer B validation and are not required to reproduce the presented results. Higher scores indicate stronger capacity. \(M\), \(S\), \(D^{P}\), and \(L^{P}\) denote macroeconomic capacity, sectoral resilience, public defence-industrial capability, and public/proxy logistics and sustainment, respectively.
\end{tablenotes}

\end{threeparttable}
\end{table}

The source hierarchy, country-specific fallbacks, vintages, and indicator-level scoring rules are reported in Online Appendix Tables 4--9.

\subsection{Unified scenario architecture}

The analysis uses one balanced baseline and five geoeconomic stress scenarios. Scenario names, symbols, stress multipliers, foreign-access assumptions, and calibration parameters are identical in the paper, online appendix, and replication files. Top-level component weights remain fixed across scenarios; stress is introduced through indicator-level transformations. Scenario analysis is a stress-testing exercise. We do not speculate about the events that might trigger these scenarios, nor do we assign probabilities to their occurrence or treat them as desirable outcomes.

\begin{table}[!htbp]
\centering
\small
\renewcommand{\arraystretch}{1.00}
\begin{threeparttable}
\caption{Overview of geoeconomic stress scenarios}
\label{tab:scenario_taxonomy}
{\linespread{0.96}\selectfont
\begin{tabularx}{\textwidth}{@{}
>{\raggedright\arraybackslash}p{0.08\textwidth}
>{\raggedright\arraybackslash}p{0.20\textwidth}
>{\raggedright\arraybackslash}X
@{}}
\toprule
\makecell[l]{Code} &
\makecell[l]{Scenario} &
\makecell[l]{Core mechanism} \\
\midrule
BL & Balanced baseline & Elevated defence demand and geopolitical tension with moderate supply-chain stress and currently expected allied availability. \\
\addlinespace[2pt]
TRS & Transfer-and-replenishment stress & Combat attrition occurs outside the sample; European effects arise through transfers, inventory drawdown, replacement demand, procurement queues, maintenance pressure, and replenishment requirements. \\
\addlinespace[2pt]
PR & Prolonged rivalry & Persistent strategic competition with gradual adaptation, managed de-risking, supplier diversification, longer-term contracting, and progressive capacity expansion. \\
\addlinespace[2pt]
SCF & Supply-chain fragmentation & Trade restrictions, critical-input disruption, technology denial, supplier concentration, transport disruption, and reduced substitutability. \\
\addlinespace[2pt]
ESR & External-support reduction & Effective access to foreign/allied production, inventories, maintenance, enabling capability, or delivery capacity declines; the principal model channel is a reduction in \(A_{c,g,s}\). \\
\addlinespace[2pt]
CS & Combined stress & Simultaneous transfer/replenishment pressure, supply-chain fragmentation, price and fiscal stress, and reduced external support. \\
\bottomrule
\end{tabularx}
}
\begin{tablenotes}[flushleft]
\footnotesize
\item Note: Under TRS, the kinetic event is external to the analysed countries; the model captures the resulting non-kinetic economic-industrial transmission into Europe. The corresponding classification is reported in Online Appendix Table 12; parameter domains, exact Low, Central, and High vectors, and indicator-level transformation rules are reported in Online Appendix Tables 13--15.
\end{tablenotes}
\end{threeparttable}
\end{table}

Low, Central, and High calibrations represent scenario intensity rather than alternative component-weighting schemes.

\section{Readiness-enabling index}
\label{sec:index}

\subsection{Aggregation}

The index combines 25 indicators into four components: macroeconomic capacity \(M\), sectoral resilience \(S\), defence-industrial capability \(D\), and logistics and sustainment \(L\). All scores lie on \([0,1]\), with higher values indicating stronger readiness-enabling conversion capacity. Within each component, applicable indicators receive equal weight. Structural non-applicability, such as absence of armed forces in Iceland, is handled by reweighting the remaining indicators rather than assigning a synthetic score.

For component \(K\in\{M,S,D,L\}\),
\begin{equation}
K_{c,s}
=
\sum_{i\in I_K}w'_{ci}N_{c,i,s},
\label{eq:component_score}
\end{equation}
where \(N_{c,i,s}\) is the normalised indicator score and \(w'_{ci}\) is the applicable within-component weight. The composite index gives equal weight to the four components in the baseline and in every stress scenario:
\begin{equation}
R_{c,s}
=
\frac{1}{4}
\left(
M_{c,s}+S_{c,s}+D_{c,s}+L_{c,s}
\right).
\label{eq:rei_one_page_public}
\end{equation}
Fixed top-level weights ensure that differences between scenarios reflect changes in the underlying indicators rather than changes in the aggregation rule.

\subsection{Normalisation, coverage, and confidence}

Indicators are transformed so that higher values always denote stronger capacity. Variables without natural bounds are normalised using baseline fifth and ninety-fifth percentiles among countries with Full coverage; highly skewed non-negative variables are log-transformed before normalisation. Lead-time indicators use a non-linear decay so that delays have progressively larger readiness costs over operationally relevant horizons. The same baseline normalisation bounds are retained in every scenario.

The scoring procedure separates the estimated indicator value from confidence in that value. Structural N/A observations are reweighted within the component. Partial observations use the same scoring equation as Full observations but carry lower confidence according to provenance, coverage quality, and vintage. Confidence does not alter the composite score, thereby avoiding mechanical advantages for countries with more extensive disclosure. Detailed transformations, confidence factors, and missing-data rules are reported in Online Appendix Tables 4--8; the baseline normalisation bounds retained in all stress states are reported in Online Appendix Table 9.

\subsection{Scenario transformations}

The calibrated shocks operate through sectoral resilience, defence-industrial capability, and logistics and sustainment. In all geoeconomic stress scenarios, the aggregation weights and baseline measurement scale remain fixed, and the macroeconomic component \(M\) is held constant. The shock magnitudes are calibrated from empirical evidence rather than selected as unconstrained hypothetical values. The calibration domains draw on production-network evidence on direct and indirect shock propagation and substitution \citep{acemoglu2012,boehm2019,bonadio2021}, geoeconomic and supply-chain evidence on concentrated dependencies and trade disruption \citep{farrell2019,baldwin2022,kancs2024supplychains}, and defence-industrial evidence on surge production, replacement times, lead times, and replenishment requirements \citep{cancian2020,kancs2025}. Online Appendix Table 13 reports the evidence-supported parameter domains, Table 14 reports the definitive Low, Central, and High calibration values, and Table 15 maps those parameters to the 25 baseline indicators.

TRS and PR increase pressure on replenishment demand, production and procurement lead times, surge capacity, and delivery. SCF changes strategic-input availability, trade frictions, substitutability, supplier concentration, and critical-input exposure. ESR lowers effective allied or foreign access and the sustainment channels that depend on it. CS activates several channels simultaneously using a directly calibrated parameter vector rather than mechanically multiplying the constituent scenarios. Low, Central, and High calibrations therefore represent scenario intensity within evidence-supported parameter ranges rather than arbitrary uncertainty bands.

\subsection{Interpretation}

The index is a comparative measure of the economic-industrial conditions that support mobilisation and sustained capability. The presented score is therefore interpreted together with its component and indicator decomposition.

Scenario vulnerability is the absolute loss relative to the balanced baseline:
\begin{equation}
V_{c,s}=R_{c,0}-R_{c,s}.
\label{eq:index_vulnerability}
\end{equation}
This quantity is distinct from the stressed level. A country can experience a relatively large absolute loss and still retain a stronger stressed score than a country that began from a lower baseline.

Within each scenario and intensity, countries receive the same calibrated shock vector; cross-country differences therefore arise from baseline structure, exposure, substitutability, industrial capacity, allied access, and sustainment. This symmetry supports comparative stress testing but abstracts from the asymmetric shocks experienced in actual crises.

\section{Empirical results}
\label{sec:results}

Throughout this section, scenario vulnerability is the absolute score loss defined in Equation~\ref{eq:theory_vulnerability},
\(V_{c,s}=R_{c,0}-R_{c,s}\). Because this is an absolute rather than proportional measure, a country with a high baseline can record a relatively large vulnerability while still retaining a higher stressed level than a lower-capacity country. Results therefore distinguish between the level of stressed capacity, the amount of capacity lost, and changes in cross-country rank. Unless otherwise stated, the discussion of simulated stress scenarios refers to the Central calibration; Low and High calibrations enter the sensitivity analysis. The discussion below uses Propositions 1--4 to organise the interpretation of country-level mechanisms and stress responses.

\subsection{How much baseline conversion capacity differs across countries}

Baseline readiness-enabling conversion capacity varies substantially across the sample. The mean estimated score is \(0.610\), with a standard deviation of \(0.083\), and the country range extends from \(0.471\) to \(0.759\). Germany records the highest baseline score (\(0.759\)), followed by France (\(0.749\)), Sweden (\(0.725\)), Italy (\(0.712\)), and the United Kingdom (\(0.708\)). At the lower end are Albania (\(0.518\)), Latvia (\(0.517\)), Iceland (\(0.486\)), North Macedonia (\(0.484\)), and Montenegro (\(0.471\)).

The component structure is at least as informative as the overall ranking. Across countries, mean scores are \(0.550\) for macroeconomic capacity \(M\), \(0.787\) for sectoral resilience \(S\), \(0.558\) for defence-industrial capability \(D\), and \(0.545\) for logistics and sustainment \(L\). The widest cross-country ranges occur in \(D\), from \(0.294\) to \(0.862\), and \(M\), from \(0.279\) to \(0.823\). By comparison, \(S\) is compressed between \(0.718\) and \(0.829\), while \(L\) ranges from \(0.354\) to \(0.657\). The baseline ranking is therefore driven primarily by differences in macroeconomic and defence-industrial conversion capacity rather than by uniformly weak sectoral resilience. The component profiles also illustrate Proposition 1's bottleneck logic: similar composite scores can conceal materially different weak links in the conversion chain.

Table~\ref{tab:baseline_country_results} reports the country-level baseline scores. The results illustrate why the component decomposition matters. Germany's baseline strength is concentrated in macroeconomic capacity and the highest \(D\) score in the sample (\(0.862\)), whereas France combines a high \(D\) score (\(0.806\)) with the strongest \(L\) score (\(0.657\)). Sweden is distinguished by the highest \(S\) score (\(0.829\)) and strong values across all four components. Denmark has the strongest \(M\) score (\(0.823\)) but a lower \(L\) score (\(0.474\)), placing it sixth overall. Norway similarly combines high \(M\) and \(S\) with weaker \(D\) and \(L\), producing an overall rank of thirteenth. At the lower end, Montenegro's score reflects particularly weak \(M\) and \(D\); North Macedonia combines the lowest \(M\) score with a middling \(L\) score; and Iceland's main weaknesses are \(D\) and \(L\). Similar aggregate scores can therefore arise from different underlying conversion constraints.

\begin{table}[htbp]
\centering
\small
\renewcommand{\arraystretch}{1.00}
\begin{threeparttable}
\caption{Baseline readiness-enabling conversion capacity by country}
\label{tab:baseline_country_results}
{\linespread{0.96}\selectfont
\begin{tabularx}{\textwidth}{@{}
>{\raggedright\arraybackslash}X
>{\centering\arraybackslash}p{0.09\textwidth}
>{\centering\arraybackslash}p{0.09\textwidth}
>{\centering\arraybackslash}p{0.09\textwidth}
>{\centering\arraybackslash}p{0.09\textwidth}
>{\centering\arraybackslash}p{0.09\textwidth}
>{\centering\arraybackslash}p{0.09\textwidth}
@{}}
\toprule
\makecell[l]{Country} &
\makecell[c]{\(M\)} &
\makecell[c]{\(S\)} &
\makecell[c]{\(D\)} &
\makecell[c]{\(L\)} &
\makecell[c]{\(R\)} &
\makecell[c]{Rank} \\
\midrule
Germany & 0.758 & 0.815 & 0.862 & 0.600 & 0.759 & 1 \\
\addlinespace[2pt]
France & 0.711 & 0.820 & 0.806 & 0.657 & 0.749 & 2 \\
\addlinespace[2pt]
Sweden & 0.702 & 0.829 & 0.753 & 0.615 & 0.725 & 3 \\
\addlinespace[2pt]
Italy & 0.646 & 0.796 & 0.798 & 0.606 & 0.712 & 4 \\
\addlinespace[2pt]
United Kingdom & 0.687 & 0.817 & 0.744 & 0.585 & 0.708 & 5 \\
\addlinespace[2pt]
Denmark & 0.823 & 0.815 & 0.654 & 0.474 & 0.691 & 6 \\
\addlinespace[2pt]
Spain & 0.596 & 0.804 & 0.715 & 0.639 & 0.689 & 7 \\
\addlinespace[2pt]
Czechia & 0.589 & 0.796 & 0.612 & 0.629 & 0.656 & 8 \\
\addlinespace[2pt]
Netherlands & 0.572 & 0.793 & 0.687 & 0.553 & 0.651 & 9 \\
\addlinespace[2pt]
Finland & 0.553 & 0.802 & 0.662 & 0.571 & 0.647 & 10 \\
\addlinespace[2pt]
Portugal & 0.654 & 0.794 & 0.567 & 0.536 & 0.638 & 11 \\
\addlinespace[2pt]
Belgium & 0.490 & 0.797 & 0.698 & 0.557 & 0.636 & 12 \\
\addlinespace[2pt]
Norway & 0.747 & 0.824 & 0.532 & 0.434 & 0.634 & 13 \\
\addlinespace[2pt]
Poland & 0.586 & 0.805 & 0.609 & 0.522 & 0.631 & 14 \\
\addlinespace[2pt]
Romania & 0.533 & 0.805 & 0.523 & 0.540 & 0.600 & 15 \\
\addlinespace[2pt]
Slovenia & 0.527 & 0.776 & 0.476 & 0.604 & 0.596 & 16 \\
\addlinespace[2pt]
Hungary & 0.359 & 0.775 & 0.505 & 0.654 & 0.573 & 17 \\
\addlinespace[2pt]
Bulgaria & 0.473 & 0.779 & 0.463 & 0.558 & 0.568 & 18 \\
\addlinespace[2pt]
Croatia & 0.523 & 0.784 & 0.428 & 0.533 & 0.567 & 19 \\
\addlinespace[2pt]
Slovakia & 0.325 & 0.782 & 0.525 & 0.616 & 0.562 & 20 \\
\addlinespace[2pt]
Estonia & 0.456 & 0.789 & 0.510 & 0.430 & 0.546 & 21 \\
\addlinespace[2pt]
Luxembourg & 0.493 & 0.758 & 0.339 & 0.561 & 0.538 & 22 \\
\addlinespace[2pt]
Lithuania & 0.467 & 0.782 & 0.442 & 0.438 & 0.532 & 23 \\
\addlinespace[2pt]
Albania & 0.523 & 0.729 & 0.321 & 0.498 & 0.518 & 24 \\
\addlinespace[2pt]
Latvia & 0.469 & 0.768 & 0.428 & 0.401 & 0.517 & 25 \\
\addlinespace[2pt]
Iceland & 0.521 & 0.761 & 0.307 & 0.354 & 0.486 & 26 \\
\addlinespace[2pt]
North Macedonia & 0.279 & 0.734 & 0.355 & 0.569 & 0.484 & 27 \\
\addlinespace[2pt]
Montenegro & 0.336 & 0.718 & 0.294 & 0.536 & 0.471 & 28 \\
\bottomrule
\end{tabularx}
}
\begin{tablenotes}[flushleft]
\footnotesize
\item Note: Scores are on the \([0,1]\) scale, with higher values indicating stronger readiness-enabling conversion capacity. Countries are ordered by the Central baseline score. The composite index gives equal weight to \(M\), \(S\), \(D\), and \(L\).
\end{tablenotes}
\end{threeparttable}
\end{table}

\subsection{How geoeconomic shocks reduce capacity}

Every simulated geoeconomic stress scenario reduces mean readiness-enabling conversion capacity, but the magnitude differs materially across scenarios. Table~\ref{tab:scenario_results_summary} reports the scenario results. The external-support-reduction scenario is the least severe in aggregate: mean \(R\) falls from \(0.610\) to \(0.588\), an absolute deterioration of \(0.022\), equivalent to \(3.6\) per cent of the baseline mean. Prolonged rivalry reduces the mean to \(0.569\), a loss of \(0.041\) or \(6.7\) per cent. Supply-chain fragmentation and transfer-and-replenishment stress generate similar average level effects in the simulation, lowering the mean to \(0.560\) and \(0.556\), respectively. Their mechanisms, however, are different, as shown below.

The combined-stress scenario is substantially more severe. Under combined stress, mean \(R\) falls to \(0.491\), an absolute loss of \(0.119\), or \(19.5\) per cent of the baseline mean. Country scores range from \(0.374\) to \(0.627\) under combined stress, compared with \(0.471\) to \(0.759\) at baseline.

\begin{table}[htbp]
\centering
\small
\setlength{\tabcolsep}{3pt}
\renewcommand{\arraystretch}{1.00}
\begin{threeparttable}
\caption{Scenario effects on readiness-enabling conversion capacity}
\label{tab:scenario_results_summary}
{\linespread{0.96}\selectfont
\begin{tabularx}{\textwidth}{@{}
>{\raggedright\arraybackslash}p{0.12\textwidth}
>{\centering\arraybackslash}p{0.11\textwidth}
>{\centering\arraybackslash}p{0.11\textwidth}
>{\centering\arraybackslash}p{0.13\textwidth}
>{\centering\arraybackslash}p{0.14\textwidth}
>{\centering\arraybackslash}p{0.11\textwidth}
>{\centering\arraybackslash}X
@{}}
\toprule
\makecell[l]{Scenario} &
\makecell[c]{Mean \(R\)} &
\makecell[c]{Mean loss} &
\makecell[c]{Loss / BL\\(\%)} &
\makecell[c]{Low--High\\range} &
\makecell[c]{\(\rho\) vs BL} &
\makecell[c]{Max rank\\move} \\
\midrule
TRS & 0.556 & 0.054 & 8.8 & 0.038 & 0.997 & 2 \\
\addlinespace[2pt]
PR & 0.569 & 0.041 & 6.7 & 0.014 & 0.998 & 1 \\
\addlinespace[2pt]
SCF & 0.560 & 0.051 & 8.3 & 0.062 & 0.998 & 2 \\
\addlinespace[2pt]
ESR & 0.588 & 0.022 & 3.6 & 0.024 & 0.999 & 1 \\
\addlinespace[2pt]
CS & 0.491 & 0.119 & 19.5 & 0.069 & 0.993 & 2 \\
\bottomrule
\end{tabularx}
}
\begin{tablenotes}[flushleft]
\footnotesize
\item Note: Mean loss is \(R_{BL}-R_s\). ``Loss / BL'' divides the mean absolute loss by the baseline sample mean. The Low--High range is the difference between the mean country score under the Low and High calibrations of each scenario. \(\rho\) is the Spearman correlation between scenario and baseline country scores.
\end{tablenotes}
\end{threeparttable}
\end{table}

Country-level losses also differ by scenario. Under TRS, the largest absolute reductions occur in France (\(0.069\)), Sweden (\(0.068\)), the United Kingdom (\(0.066\)), Italy (\(0.066\)), and Germany (\(0.066\)). PR produces almost the same ordering, with the largest losses again concentrated in France, Sweden, the United Kingdom, Italy, and Germany. This does not imply greater preparedness among lower-baseline countries; it reflects, in part, the use of an absolute loss measure: countries with larger defence-industrial capacity have more measured capacity exposed to demand, production, and replenishment shocks.

SCF produces a different pattern. Iceland records the largest central absolute loss (\(0.059\)), followed by Bulgaria, Slovakia, Lithuania, and Croatia at approximately \(0.052\). Germany, Norway, Estonia, the United Kingdom, and Sweden are among the countries with the smaller SCF losses. ESR is narrower in cross-country magnitude, with Iceland again recording the largest loss (\(0.029\)); most other countries fall close to \(0.020\)--\(0.023\). Under combined stress, the largest absolute losses occur in France, Sweden, Italy, Germany, and Spain, at approximately \(0.131\)--\(0.134\).

The main empirical implication is that strategic stress changes capacity levels much more than it changes the broad cross-country ordering. Even under combined stress, the Spearman correlation with baseline scores remains \(0.993\), and the maximum rank movement is two places. Germany and France remain first and second under every simulated scenario. High-baseline countries can nevertheless record large absolute losses because more measured capacity is exposed to production, replenishment, and sustainment shocks. Baseline strength and absolute vulnerability should therefore be read together rather than as competing rankings.

\subsection{Why countries respond differently}

The scenario engine is designed so that different shocks act on different conversion mechanisms. Because \(M\) is held constant within the stress transformations, central vulnerability can be written as:
\begin{equation}
V_{c,s}
=
\frac{1}{4}
\left[
(S_{c,0}-S_{c,s})
+
(D_{c,0}-D_{c,s})
+
(L_{c,0}-L_{c,s})
\right].
\label{eq:results_component_decomposition}
\end{equation}
Table~\ref{tab:scenario_component_decomposition} decomposes the mean deterioration.

\begin{table}[htbp]
\centering
\small
\setlength{\tabcolsep}{3pt}
\renewcommand{\arraystretch}{1.00}
\begin{threeparttable}
\caption{Component decomposition of scenario vulnerability}
\label{tab:scenario_component_decomposition}
{\linespread{0.96}\selectfont
\begin{tabularx}{\textwidth}{@{}
>{\raggedright\arraybackslash}p{0.10\textwidth}
>{\centering\arraybackslash}p{0.15\textwidth}
>{\centering\arraybackslash}p{0.12\textwidth}
>{\centering\arraybackslash}p{0.12\textwidth}
>{\centering\arraybackslash}p{0.12\textwidth}
>{\raggedright\arraybackslash}X
@{}}
\toprule
\makecell[l]{Scenario} &
\makecell[c]{Mean\\vulnerability} &
\makecell[c]{\(S\) share\\(\%)} &
\makecell[c]{\(D\) share\\(\%)} &
\makecell[c]{\(L\) share\\(\%)} &
\makecell[l]{Dominant component\\across countries} \\
\midrule
TRS & 0.054 & 0.0 & 65.7 & 34.3 & D: 25; L: 3 \\
\addlinespace[2pt]
PR & 0.041 & 0.0 & 65.0 & 35.0 & D: 24; L: 4 \\
\addlinespace[2pt]
SCF & 0.051 & 59.0 & 21.4 & 19.6 & S: 28 \\
\addlinespace[2pt]
ESR & 0.022 & 0.0 & 49.6 & 50.4 & D: 12; L: 16 \\
\addlinespace[2pt]
CS & 0.119 & 33.7 & 43.3 & 23.0 & D: 22; S: 5; L: 1 \\
\bottomrule
\end{tabularx}
}
\begin{tablenotes}[flushleft]
\footnotesize
\item Note: Component shares are the sample-average shares of total country-level deterioration. \(M\) is unchanged by design in all stress scenarios. ``Dominant component'' reports the number of countries for which the named component contributes the largest share of scenario deterioration.
\end{tablenotes}
\end{threeparttable}
\end{table}

TRS and PR are predominantly defence-industrial shocks. On average, \(D\) accounts for \(65.7\) per cent of TRS deterioration and \(65.0\) per cent of PR deterioration, with \(L\) accounting for the remainder. \(D\) is the dominant component in 25 of 28 countries under TRS and 24 under PR. At indicator level, the largest mean TRS contributions come from baseline domestic production \(D1.2\) (22.7 per cent of total deterioration), procurement activity and demand credibility \(D1.6\) (19.1 per cent), replenishment speed \(L1.5\) (16.7 per cent), and production/procurement lead time \(D1.4\) (12.0 per cent). PR has a similar profile: \(D1.2\) and \(D1.6\) jointly account for just over half of the mean deterioration, while \(L1.5\) contributes 17.6 per cent. The similarity of TRS and PR shows that the two scenarios stress closely related replenishment mechanisms despite their different horizons; this becomes important for evaluating Proposition 2 below.

SCF is qualitatively different. Sectoral resilience \(S\) contributes 59.0 per cent of the mean deterioration and is the dominant component in all 28 countries. The single largest indicator contribution is input and supplier substitutability \(S1.4\), accounting for 25.9 per cent of mean SCF deterioration. Indirect upstream exposure \(S1.2\) contributes 11.7 per cent, critical-material and technology dependence \(S1.5\) 9.3 per cent, and supplier concentration \(S1.3\) 6.3 per cent. The largest non-\(S\) contribution is replenishment speed \(L1.5\), at 15.2 per cent. The contrast with TRS and PR supports Proposition 2: supply-chain fragmentation exposes a distinct set of upstream and substitutability constraints.

ESR is concentrated in access and sustainment channels. The mean deterioration is split almost evenly between \(D\) (49.6 per cent) and \(L\) (50.4 per cent). Effective allied or foreign supply access \(D1.8\) alone contributes 45.8 per cent of the mean loss, while replenishment speed \(L1.5\) contributes 35.8 per cent. The concentration of the ESR loss in allied-access and replenishment channels is consistent with Proposition 3: foreign supply expands effective capacity in the baseline but creates contingent vulnerability when access and delivery deteriorate.

Under combined stress, several geoeconomic shocks occur simultaneously and affect all adjustment channels. \(D\) accounts for 43.3 per cent of mean deterioration, \(S\) for 33.7 per cent, and \(L\) for 23.0 per cent. \(D\) is the dominant component in 22 countries, \(S\) in five, and \(L\) in one. No single indicator dominates the combined result. The largest contributions are \(S1.4\) substitutability (13.7 per cent), \(D1.2\) baseline domestic production (12.1 per cent), \(D1.8\) effective allied access (10.5 per cent), \(L1.5\) replenishment speed (9.4 per cent), and \(D1.6\) procurement activity (8.6 per cent). CS therefore operates as a multi-channel stress rather than as a simple rescaling of one component.

The country profiles reinforce this mechanism-based interpretation. Germany and France retain the highest stressed scores because they begin with broad strength across \(M\), \(D\), and \(L\), even though their absolute TRS and CS losses are comparatively large. Iceland's relative weakness shifts by scenario: its baseline constraints are concentrated in \(D\) and \(L\), while SCF and ESR expose additional dependence on supply-chain and access channels. Norway provides the converse example: its high \(M\) and \(S\) support the baseline score, but lower \(D\) and \(L\) constrain replenishment-oriented performance. The same aggregate baseline score therefore cannot be interpreted without its component structure.

\subsection{How robust and distinct the scenarios are}

Scenario validation tests implementation integrity, calibration sensitivity, rank stability, channel importance, and scenario discrimination. All nine checks pass. Scenario scores remain within \([0,1]\), structural N/A observations are preserved, baseline indicator and component scores reconcile, Low--Central--High intensity is monotonic, indicators specified as unchanged remain invariant, component and top-level weights sum correctly, and the workbook contains no formula errors. Independent recomputation of the scenario results from the embedded baseline inputs, bounds, calibration, and transformation rules yields a maximum discrepancy of \(1.11\times10^{-16}\).

Calibration uncertainty affects levels more than ranks. As Table~\ref{tab:scenario_robustness} shows, the widest interval in the mean country score is under CS, from \(0.531\) at Low intensity to \(0.461\) at High intensity, followed by SCF (\(0.582\) to \(0.521\)), TRS (\(0.574\) to \(0.536\)), ESR (\(0.600\) to \(0.576\)), and PR (\(0.575\) to \(0.561\)). Yet scenario rank correlations with the baseline all exceed \(0.993\), and the maximum scenario rank shift is only two places. Even under the High CS calibration, the maximum observed rank movement is three places. The principal findings are therefore not driven by large rank reversals across plausible shock intensities.

\begin{table}[htbp]
\centering
\small
\setlength{\tabcolsep}{3pt}
\renewcommand{\arraystretch}{1.00}
\begin{threeparttable}
\caption{Scenario robustness and channel sensitivity}
\label{tab:scenario_robustness}
{\linespread{0.96}\selectfont
\begin{tabularx}{\textwidth}{@{}
>{\raggedright\arraybackslash}p{0.08\textwidth}
>{\centering\arraybackslash}p{0.14\textwidth}
>{\centering\arraybackslash}p{0.08\textwidth}
>{\centering\arraybackslash}p{0.12\textwidth}
>{\raggedright\arraybackslash}p{0.15\textwidth}
>{\raggedright\arraybackslash}X
@{}}
\toprule
\makecell[l]{Scenario} &
\makecell[c]{Mean \(R\)\\\textnormal{[Low, High]}} &
\makecell[c]{\(\rho\) vs BL} &
\makecell[c]{Max rank\\move} &
\makecell[l]{Largest within-\\range effect} &
\makecell[l]{Largest leave-one-channel-out\\recovery} \\
\midrule
TRS & [0.574, 0.536] & 0.997 & 2 & \(\theta^{S*}\) (0.0136) & Replenishment demand (0.0303) \\
\addlinespace[2pt]
PR & [0.575, 0.561] & 0.998 & 1 & \(\theta^{S*}\) (0.0091) & Replenishment demand (0.0274) \\
\addlinespace[2pt]
SCF & [0.582, 0.521] & 0.998 & 2 & \(\theta^{M}\) (0.0222) & Substitutability (0.0159) \\
\addlinespace[2pt]
ESR & [0.600, 0.576] & 0.999 & 1 & \(\theta^{A}\) (0.0236) & Allied access (0.0211) \\
\addlinespace[2pt]
CS & [0.531, 0.461] & 0.993 & 2 & \(\theta^{M}\) (0.0197) & Replenishment demand (0.0317) \\
\bottomrule
\end{tabularx}
}
\begin{tablenotes}[flushleft]
\footnotesize
\item Note: The second column reports the mean country score under the [Low-intensity, High-intensity] calibration for each scenario. Parameter effects report the mean absolute country effect between that parameter's calibrated Low and High values around the central specification. Leave-one-channel-out values report the mean recovery in \(R\) when the named stress channel is neutralised. The two diagnostics answer different questions and need not identify the same parameter as most influential.
\end{tablenotes}
\end{threeparttable}
\end{table}

Within the calibrated ranges, TRS is most sensitive to replenishment demand \(\theta^{S*}\), allied access \(\theta^A\), and lead time \(\theta^L\). SCF is most sensitive to import availability \(\theta^M\), substitutability \(\theta^\sigma\), and trade or transport friction \(\theta^\tau\). ESR is driven by allied access \(\theta^A\). CS has the broadest sensitivity profile: import availability has the largest within-range effect, followed by substitutability and allied access. In the leave-one-channel-out exercise, however, replenishment demand produces the largest mean CS recovery (\(0.0317\)), followed by substitutability (\(0.0204\)), import availability (\(0.0179\)), and allied access (\(0.0169\)). This difference reflects the distinction between variation within an evidence-supported calibration interval and complete neutralisation of a channel.

The scenarios also differ in the cross-country pattern of vulnerability they generate. TRS and PR vulnerability vectors are almost identical in rank terms (\(\rho=0.999\)), showing that the two scenarios do not generate distinct country vulnerability rankings despite their different horizons. By contrast, SCF is negatively correlated with TRS (\(\rho=-0.247\)) and PR (\(\rho=-0.259\)), and ESR is only weakly associated with the replenishment-oriented scenarios. Proposition 2 is therefore supported unevenly: supply-chain fragmentation and external-support reduction reveal distinct channels, whereas TRS and PR largely identify the same replenishment vulnerability. CS remains strongly correlated with TRS and PR (\(\rho=0.992\) and \(0.991\)), indicating that replenishment-related constraints remain important when several channels are activated simultaneously.

The combined-stress scenario also passes the pre-specified dominance diagnostic. For all 28 countries, CS vulnerability is at least as large as the maximum of TRS, SCF, and ESR vulnerability; there are no dominance exceptions. This is consistent with Proposition 4. The outcome is not mechanically imposed because CS is directly calibrated rather than constructed as a product of the constituent scenarios.

One specification issue remains visible in the validation results. The surge-activation parameter \(\phi\) enters \(D1.3\) in every stress state. Consequently, SCF and ESR produce a small \(D1.3\) effect even though \(\theta^S=1\) in those scenarios. Neutralising this spillover raises mean \(R\) by only \(0.00095\) in SCF and \(0.00081\) in ESR, with rank correlations of \(0.998\) and \(1.000\), respectively, relative to the full specification. The effect is therefore quantitatively small, but it weakens the interpretation of ESR as a fully isolated external-access shock. The issue is retained and documented in the online appendix.

\section{External validation}
\label{sec:external_validation}

External validation assesses whether the estimated readiness index behaves as expected relative to independent measures of industrial capability, innovation, logistics, and economic integration. The exercise is a test of our framework's coherence rather than a search for a benchmark that reproduces the index. No external series is treated as a criterion measure of readiness-enabling conversion capacity, because the conceptual framework presented in Section~\ref{sec:readiness_theory} combines mechanisms that existing indicators measure only partially.

\subsection{Validation design}

The validation design follows three principles. First, the comparison is conducted against benchmarks that are independent of the readiness index and are not used in its construction. Second, Spearman rank correlation is the primary statistic because the paper is concerned principally with cross-country ordering and because several external measures are ranks or non-linear proxies. Pearson correlation is reported as a supplementary measure of linear association. Third, interpretation emphasises effect size, expected sign, sample size, and conceptual relevance rather than statistical significance as a pass--fail criterion.

\begin{sidewaystable}[p]
\centering
\small
\setlength{\tabcolsep}{3pt}
\renewcommand{\arraystretch}{1.00}
\begin{threeparttable}
\caption{External validation of the readiness-enabling conversion measure}
\label{tab:external_validation}
{\linespread{0.96}\selectfont
\begin{tabularx}{0.95\textheight}{@{}
>{\raggedright\arraybackslash}p{0.05\textheight}
>{\raggedright\arraybackslash}p{0.32\textheight}
>{\centering\arraybackslash}p{0.08\textheight}
>{\centering\arraybackslash}p{0.08\textheight}
>{\centering\arraybackslash}p{0.08\textheight}
>{\centering\arraybackslash}p{0.08\textheight}
>{\raggedright\arraybackslash}X
@{}}
\toprule
\makecell[l]{Target} &
\makecell[l]{External benchmark} &
\makecell[c]{\(N\)} &
\makecell[c]{Pearson\\\(r\)} &
\makecell[c]{Spearman\\\(\rho\)} &
\makecell[c]{Sign\\met} &
\makecell[l]{Interpretation} \\
\midrule
\(L\) & World Bank LPI overall & 28 & 0.096 & 0.094 & Yes & Weak / no evidence \\
\addlinespace[2pt]
\(D\) & UNIDO CIP global rank & 28 & -0.844 & -0.911 & Yes & Strong \\
\addlinespace[2pt]
\(D\) & UNIDO manufacturing value added per capita & 28 & 0.596 & 0.587 & Yes & Meaningful \\
\addlinespace[2pt]
\(D\) & UNIDO medium-high/high-technology manufacturing share & 28 & 0.682 & 0.617 & Yes & Strong \\
\addlinespace[2pt]
\(D\) & SIPRI Top 100 arms revenues & 28 & 0.532 & 0.658 & Yes & Strong \\
\addlinespace[2pt]
\(D\) & SIPRI Top 100 firm presence & 28 & 0.674 & 0.686 & Yes & Strong \\
\addlinespace[2pt]
\(D\) & SIPRI Top 100 revenues, positive subset & 9 & 0.461 & 0.683 & Yes & Strong \\
\addlinespace[2pt]
\(D\) & WIPO Global Innovation Index & 28 & 0.696 & 0.683 & Yes & Strong \\
\addlinespace[2pt]
\(D\) & Manufacturing value added, \% of GDP & 28 & 0.397 & 0.322 & Yes & Modest \\
\addlinespace[2pt]
\(S\) & EC Single Market goods-trade integration & 28 & -0.148 & -0.255 & No & Modest, opposite sign \\
\addlinespace[2pt]
\(R\) & World Bank LPI overall & 28 & 0.704 & 0.703 & Yes & Strong \\
\addlinespace[2pt]
\(R\) & UNIDO CIP global rank & 28 & -0.817 & -0.880 & Yes & Strong \\
\addlinespace[2pt]
\(R\) & WIPO Global Innovation Index & 28 & 0.709 & 0.698 & Yes & Strong \\
\bottomrule
\end{tabularx}
}
\begin{tablenotes}[flushleft]
\footnotesize
\item Note: Spearman rank correlation is the primary statistic. CIP is rank-coded with 1 denoting the strongest industrial performance, so the expected sign is negative. SIPRI all-country revenues retain true zeros, which denote the absence of a nationally headquartered 2024 SIPRI Top 100 firm rather than zero domestic arms production. The positive-revenue SIPRI test is supplementary and uses \(N=9\). The EC measure captures Single Market goods-trade integration, not external strategic vulnerability.
\end{tablenotes}
\end{threeparttable}
\end{sidewaystable}

The principal benchmarks are the World Bank Logistics Performance Index (LPI), the UNIDO Competitive Industrial Performance (CIP) rank, UNIDO manufacturing-depth and technology-intensity indicators, SIPRI Top 100 arms-industry revenues and firm presence, the WIPO Global Innovation Index (GII), World Bank manufacturing value added as a share of GDP, and the European Commission Single Market goods-trade integration measure \citep{worldbank2023lpi,unido2023cip,unido2024stats,sipri2025armsindustry,wipo2025gii,worldbank2024mva,ec2024singlemarket}. For interpretation, absolute Spearman correlations below 0.20 are classified as weak or no evidence, values from 0.20 to below 0.40 as modest, values from 0.40 to below 0.60 as meaningful, and values of 0.60 or above as strong.

Table~\ref{tab:external_validation} reports the complete external-validation set. All tests use the full 28-country sample except the supplementary SIPRI positive-revenue test, which conditions on the nine countries with a nationally attributed Top 100 arms-industry revenue observation greater than zero.

\subsection{Defence-industrial convergent validity}

The strongest external evidence concerns the defence-industrial component \(D\). Every defence-industrial benchmark has the expected sign. The association with the UNIDO CIP rank is particularly strong, with \(\rho=-0.911\). Because lower CIP rank numbers indicate stronger industrial performance, the negative sign is the expected one. The result indicates that countries scoring highly on the defence-industrial component also tend to occupy stronger positions in an independent measure of industrial competitiveness.

\(D\) is meaningfully associated with manufacturing value added per capita (\(\rho=0.587\)) and strongly associated with the medium-high and high-technology share of manufacturing value added (\(\rho=0.617\)). It is also strongly associated with the Global Innovation Index (\(\rho=0.683\)). These relationships are consistent with the interpretation of \(D\) as measuring more than the presence of defence firms: industrial depth, technological intensity, and innovation capacity all matter for the ability to expand and sustain complex defence production.

The defence-specific tests point in the same direction. The correlation between \(D\) and nationally attributed SIPRI Top 100 arms revenues is \(\rho=0.658\) for the full 28-country sample, while the binary presence of at least one nationally headquartered Top 100 firm yields \(\rho=0.686\). Conditioning on the nine countries with positive nationally attributed Top 100 revenues gives \(\rho=0.683\). The positive-subset test is supplementary because of the small sample size, but its direction and magnitude are consistent with the full-sample results.

By contrast, the association with manufacturing value added as a share of GDP is only modest (\(\rho=0.322\)). We interpret this as informative rather than contradictory. A high manufacturing share is not equivalent to defence-industrial conversion capacity: sectoral composition, technology intensity, supplier depth, procurement activity, surge potential, qualification constraints, and access to allied supply all matter. Taken together, the \(D\) results provide strong convergent evidence while also showing that the component is not a relabelled measure of aggregate manufacturing weight.

\subsection{Logistics and sectoral resilience}

External validation is unresolved for \(L\) and \(S\), for which closely matched external benchmarks are scarce. The logistics-and-sustainment component \(L\) has only a weak association with the World Bank LPI (\(\rho=0.094\)), providing no evidence of convergent validity. Because the LPI measures commercial logistics rather than defence sustainment, this result does not test several mechanisms contained in \(L\), including maintenance, spare-parts support, pre-positioning, replenishment speed, and replacement capacity. The appropriate conclusion is therefore that external validation of \(L\) remains open.

The sectoral-resilience comparison is also non-confirmatory. The correlation between \(S\) and the European Commission Single Market goods-trade integration measure is modest and negative (\(\rho=-0.255\)), contrary to the expected positive sign. The benchmark captures market integration rather than external strategic vulnerability, so it is an imperfect comparator; nevertheless, it does not validate \(S\). External validation of the sectoral-resilience component therefore also remains unresolved.

Weak or oppositely signed relationships are reported as non-confirmatory rather than reinterpreted as supportive evidence.

\subsection{Robustness of the composite index}

The composite index \(R\) exhibits strong external coherence across three broad and independent benchmarks. Its Spearman correlation with the World Bank LPI is \(0.703\), with the UNIDO CIP rank it is \(-0.880\), and with the Global Innovation Index it is \(0.698\). The corresponding Pearson correlations are \(0.704\), \(-0.817\), and \(0.709\), respectively.

None of the three benchmarks is used in constructing \(R\). Their associations with the composite index therefore provide evidence of external coherence across industrial, innovation, and general logistics capability, while the weak direct \(L\)-LPI relationship cautions against treating that coherence as component-level validation of defence sustainment.

\section{Discussion}
\label{sec:discussion}

\subsection{Interpretation of the empirical findings}

Three implications emerge from the results. First, the persistence of the cross-country ordering under common shocks indicates that structural conversion capacity matters more for relative position than short-run scenario exposure. Combined stress lowers measured capacity substantially, yet countries with broad macroeconomic, industrial, and sustainment strengths remain comparatively strong. The contribution of stress testing is therefore not primarily to generate a new ranking, but to show how much capacity is lost and which mechanism accounts for that loss.

Second, conceptually distinct scenarios do not always produce empirically distinct vulnerability rankings. TRS and PR are almost indistinguishable in rank terms, suggesting that their different horizons and calibration assumptions act on a common replenishment bottleneck. SCF, by contrast, produces a markedly different pattern centred on substitutability and upstream exposure, while ESR isolates the contingency created by foreign access. This finding is informative: scenario differentiation should be judged by the mechanisms and country patterns it reveals, not by scenario labels alone.

Third, the validation evidence is uneven in a way that helps define where the index is mature and where further measurement work is needed. The defence-industrial component has strong convergent relationships with independent industrial, technological, innovation, and defence-firm measures. External validation of sectoral resilience and defence-oriented logistics remains unresolved because the available comparators either measure adjacent concepts or fail to correlate with the proposed component. The composite index is externally coherent, but that coherence should not be read as validating every component equally.

\subsection{Policy implications}

A central policy implication is to complement expenditure targets with measures of conversion performance. Relevant metrics include replacement time after inventory depletion, qualified substitute suppliers, the share of procurement covered by credible multi-year contracts, maintenance and spare-parts capacity under higher utilisation, crisis throughput on transport corridors, and dependence on concentrated external inputs. These measures identify the bottlenecks that determine whether additional expenditure can be converted into regeneration and sustainment within the relevant time horizon.

The results also underline the value of coordinated procurement and industrial policy, especially when coordination relaxes a binding constraint. Common procurement can create credible demand, standardise requirements, and support longer production runs. It can also intensify queues when governments compete for the same constrained energetic materials, electronics, testing facilities, or skilled labour. Similarly, investment in assembly capacity has limited value if upstream inputs, certification, maintenance, or transport remain binding. The component and indicator decompositions provide a way to distinguish these cases.

Allied access is part of readiness rather than a separate dimension. Foreign supply can materially increase effective replenishment capacity, but its value depends on reliable contractual access, export and licensing conditions, transport, delivery schedules, supplier-country concentration, and competing allied demand. Policy assessments should therefore consider both domestic industrial capacity and the reliability and timing of external supply.

\subsection{Limitations and avenues for future research}

The scenario exercise is deliberately comparative rather than predictive. Within a given scenario and intensity, all countries are exposed to the same calibrated shock vector; the shock vector differs across scenarios and intensity levels. This symmetry makes it possible to attribute cross-country differences to baseline structures and exposures, but it abstracts from the strongly asymmetric way real crises affect countries. Actual shocks differ in timing, geography, physical destruction, battlefield attrition, infrastructure damage, political decisions, and access to external support. Ukraine after Russia's full-scale invasion illustrates the magnitude of such asymmetry: destruction of productive capacity and infrastructure, equipment losses, disrupted logistics, and simultaneous military and civilian regeneration demands are qualitatively more severe than the common non-kinetic shocks imposed in this study. The simulated losses should not be interpreted as upper bounds on wartime deterioration.

Data limitations also remain important. Open-source and licensable data do not fully reveal stockpiles, maintenance backlogs, supplier contracts, surge capacity, or the readiness state of specific production lines. Input-output data are often too aggregated to identify specific defence components, and some defence-industrial indicators require mapped or proxy observations. The confidence framework records these limitations without allowing disclosure intensity to determine the Central score.

Future research should use more granular longitudinal data on procurement, production, inventories, maintenance, deliveries, and supply-chain disruptions to estimate how economic-industrial conditions affect realised mobilisation and replenishment outcomes. Such evidence could also inform data-driven component weights. Extending the present comparative statics to asymmetric country-specific shocks, stock depletion, endogenous investment, cross-border spillovers, and dynamic adaptation would provide a stronger account of how conversion capacity evolves during prolonged strategic stress.

\section{Conclusion}
\label{sec:conclusion}

European defence readiness in a protracted conflict depends not only on force structure and defence expenditure, but also on the capacity to convert economic, industrial, and logistical resources into replenishment and sustained military capability. This study operationalises that intermediate mechanism as readiness-enabling conversion capacity and measures it across 28 European NATO Allies using macroeconomic capacity, sectoral resilience, defence-industrial capability, and logistics and sustainment.

The empirical results show substantial cross-country variation at baseline and sizeable deterioration under strategic stress. Combined stress reduces average measured capacity by 19.5 per cent relative to baseline, while the single-channel scenarios expose different vulnerabilities. Replenishment-oriented scenarios are driven mainly by defence-industrial and logistics constraints; supply-chain fragmentation is driven primarily by sectoral exposure and substitutability; and external-support reduction highlights the importance of reliable allied access. Country rankings remain comparatively stable, indicating that the simulated geoeconomic shocks alter capacity levels far more than they alter the cross-country ordering.

External validation is strongest for the defence-industrial component and the composite index, while validation of sectoral resilience and defence-oriented logistics remains unresolved. The component decomposition nevertheless reveals why similar aggregate scores can conceal different bottlenecks and why the same country can be resilient to one form of stress while remaining exposed to another.

The central policy implication is that defence resources should be assessed together with conversion performance---production lead times, supplier substitutability, replenishment speed, maintenance capacity, logistics throughput, and the reliability of allied supply. The symmetric stress tests used here provide a common comparative benchmark. Extending the framework to asymmetric shocks, dynamic investment responses, stockpile depletion, and cross-border production networks would provide a fuller account of how European defence capability can be sustained during protracted conflict.

\end{document}